# A Semantic Segmentation Network Based Real-Time Computer-Aided Diagnosis System for Hydatidiform Mole Hydrops Lesion Recognition in Microscopic View

Chengze Zhu, Pingge Hu, *Student Member, IEEE*, Xianxu Zeng, Xingtong Wang, *Student Member, IEEE*, Zehua Ji and Li Shi

*Abstract*—As a disease with malignant potential, hydatidiform mole (HM) is one of the most common gestational trophoblastic diseases. For pathologists, the HM section of hydrops lesions is an important basis for diagnosis. In pathology departments, the diverse microscopic manifestations of HM lesions and the limited view under the microscope means that physicians with extensive diagnostic experience are required to prevent missed diagnosis and misdiagnosis. Feature extraction can significantly improve the accuracy and speed of the diagnostic process. As a remarkable diagnosis assisting technology, computer-aided diagnosis (CAD) has been widely used in clinical practice. We constructed a deep learning based CAD system to identify HM hydrops lesions in the microscopic view in real time. The system consists of three modules; the image mosaic module and edge extension module process the image to improve the outcome of the hydrops lesion recognition module, which adopts a semantic segmentation network, our novel compound loss function, and a stepwise training function in order to achieve the best performance in identifying hydrops lesions. We evaluated our system using an HM hydrops dataset. Experiments show that our system is able to respond in real time and correctly display the entire microscopic view with accurately labeled HM hydrops lesions.

*Index Terms*—Medical diagnostic imaging, Artificial intelligence, Engineering in medicine and biology, Pathology

## I. INTRODUCTION

In recent years, computer-aided diagnosis (CAD) has been widely used in anomaly detection and the differential diagnosis of medical images obtained by different imaging methods. The CAD system is able to process the input images in a short time, mark the parts that might be suspicious, and help physicians establish correct diagnoses. Hydatidiform mole (HM) is a gestational trophoblastic disease (GTD) that has a certain probability of growing into invasive HM and choriocarcinoma. Fetuses developed alongside HM are unlikely to survive and often have malformations, which may lead to complications or miscarriage [1]-[3]. HM before 12 weeks of gestation is often morphologically confused with other diseases or not detected due to its incomplete development [4][5]. Such misdiagnoses and missed diagnoses delay treatment. Microscopic histopathologic diagnosis, the gold standard for HM, mainly involves observing the villi features in the tissue sections, since the villi features of HM mostly manifest as villi trophoblast proliferation and interstitial hydrops[3][6]. Pathologists must spend a great deal of time studying various tissue sections under the microscope on a daily basis, and the detection efficiency is relatively low. Therefore, there is an urgent need for an HM CAD system that can improve diagnostic accuracy, reduce the rate of missed diagnosis and misdiagnosis, and lessen the workload of physicians.

The traditional CAD framework consists of feature extraction, feature selection, and classification [7], and the extracted effective features are highly related to the diagnostic task [8]-[10]. In general, the diagnostic effect of a traditional CAD system is highly dependent on hand-crafted features and requires a series of complex image processing steps to obtain a satisfactory result [11][12]. Therefore, it is usually difficult to design the algorithm framework of a traditional CAD system. In recent years, deep learning has proven effective in improving the diagnostic accuracy and efficiency of pathologists in many medical imaging recognition tasks, and has been successfully applied in lung [13] and head [14] CT images, fundus images [15], cardiomegaly X-ray images [16] and breast ultrasound images[8]. Deep learning models have also been used to detect lesions in pathology images obtained from breast cancer pathological sections [17] [18], urothelial carcinoma pathological sections[19] and colon cancer pathological sections[20][21]. P. Pal et al. [1] classified path-ological sections of hydatidiform mole villi into normal, PHM, or CHM categories based on characteristics of hydatidiform mole using three fully connected networks. However, this method can only extract the superficial features of the image and classify the whole image. The algorithm does not have the capability to locate and segment the lesion from the whole image. At present,

C. Zhu, P. Hu, X. Wang, Z. Ji and L. Shi are with the Department of Automation, Tsinghua University, Beijing, 100084 P.R.China. (e-mail: zhucz18@tsinghua.org.cn;hpg18@mails.tsinghua.edu.cn;xingtong21@mails.tsinghua.edu.cn; jizh18@mails.tsinghua.edu.cn; shilits@mail.tsinghua.edu.cn)..

X. Zeng is with the Department of Pathology, the Third Affiliated Hospital of Zhengzhou University, Zhengzhou, Henan, 450052 P.R.China. (e-mail: xianxu77@163.com).

C. Zhu and P. Hu are with equal contribution.

most projects based on deep learning are in the research stage, and few can be directly used in clinical diagnosis. There is a lack of real-time CAD systems for microscopic section images that can effectively apply network models to real-world clinical data. In 2019, an article in Nature on Google's smart microscope mentioned a method for identifying lesions in prostate pathological sections using image classification networks [22]. This method is capable of scan-ning large section images, but it is not suitable for identifying lesions in HM section images since these images have low resolution. When processing images with low resolution, the classification network is not able to obtain results with clear enough edges. Thus, we will not discuss this approach.

In this paper, a real-time CAD system for identifying HM hydrops lesions in the microscopic view is introduced. Different from the single lesion recognition model, this CAD system includes an image mosaic module, an edge extension module, and a compound loss function based algorithm module for HM hydrops lesion recognition. The CAD system can display the microscopic view with labeled hydrops lesions, the historical view mosaic image, and the historical view hydrops lesion image in real time. Thus, it can help pathologists identify and diagnose HM hydrops lesions with greater speed and higher accuracy. Not only does this system provide a solution to improve the process of HM diagnosis, but it is also generalizable and has great potential to be adapted for use in other areas of clinical practice.

## II. THE PROPOSED METHOD

### A. Dataset

The data used in this study were from *the Third Affiliated Hospital of Zhengzhou University*. A total of 157 HM biopsies from 59 patients were selected, and all HM biopsies were scanned by Motic's biopsy scanner as the main subsequent biopsy data set. The data usage was approved by *the Third Affiliated Hospital of Zhengzhou University Ethic Committee*. To reduce the class imbalance of the dataset, 3078 pieces of images with hydrops areas and 3724 pieces without hydrops areas were cleaned as training and validation dataset. As a result, 96 pieces of images without hydrops areas were randomly selected with probability 0.025. A total of 3174 pieces were split into train and valid at a ratio of 9:1. To evaluate the model performance on a new section, test images are from different subjects of training and validation dataset. The testing dataset finally obtained 330 images in the same cleaning way.

The data set used for the subsequent HM hydrops lesion rec ognition model was obtained from patients ranging in age from 25 to 38, which is the primary age of pregnancy for women, a nd the length of menolipsis ranged from 30 to 92 days. Most se ctions were from patients with complete or partial HM, while t he rest were from patients with several other diseases that are o ften confused with HM.

With the help of doctors, we assigned three different labels (hydrops, hyperplasia, and villus) to the scanned section images. All the labeling results were reviewed and approved by pathologists. Section images and labeling files were obtained by using the *Motic Digital Section Assistant system*. After data conversion and image processing, the labeled diagrams of HM sections and hydrops lesions of HM section were obtained. Two examples are shown in Figure 1.

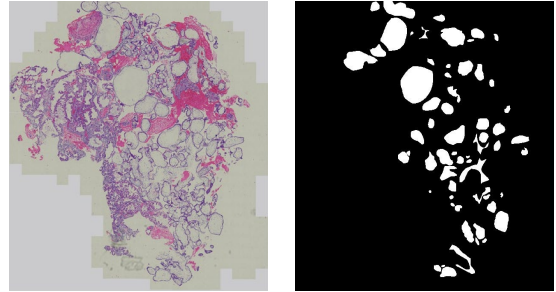

Figure 1. Scan and label mask of HM. Left: HM section scan under microscope; Right: Label mask of HM.

### B. CAD System Structure

The developed CAD system included a microscope, microscope camera, computer host, display screen, and other hardware, as well as an image mosaic module, edge extension module, hydrops lesion recognition module, and other software modules. The structure diagram of this real-time CAD system for HM hydrops lesion recognition in the microscopic view is shown in Figure 2.

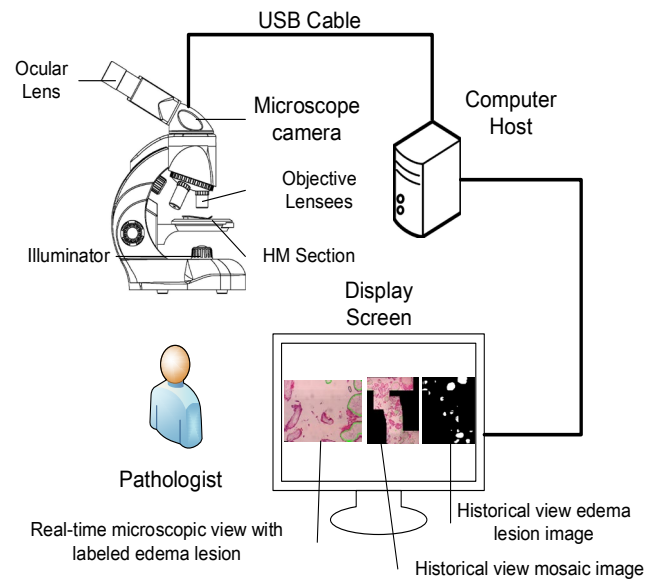

Figure 2. Structure diagram of the real-time CAD system for HM edema lesion in microscopic field.

The microscope used in this study was the CX31 binocular microscope (objective multiple 10×, eyepiece multiple 10×, mi croscope camera model G1UD05C) from Olympus Corporatio n. The microscope and camera were provided by the Departme nt of Pathology of *the Third Affiliated Hospital of Zhengzhou University*. The selected equipment and parameters are conven tional and non-specified in order to lower the difficulties for ot hers to implement our system.

The microscope cameras used by the Department of Pathology of *the Third Affiliated Hospital of Zhengzhou*

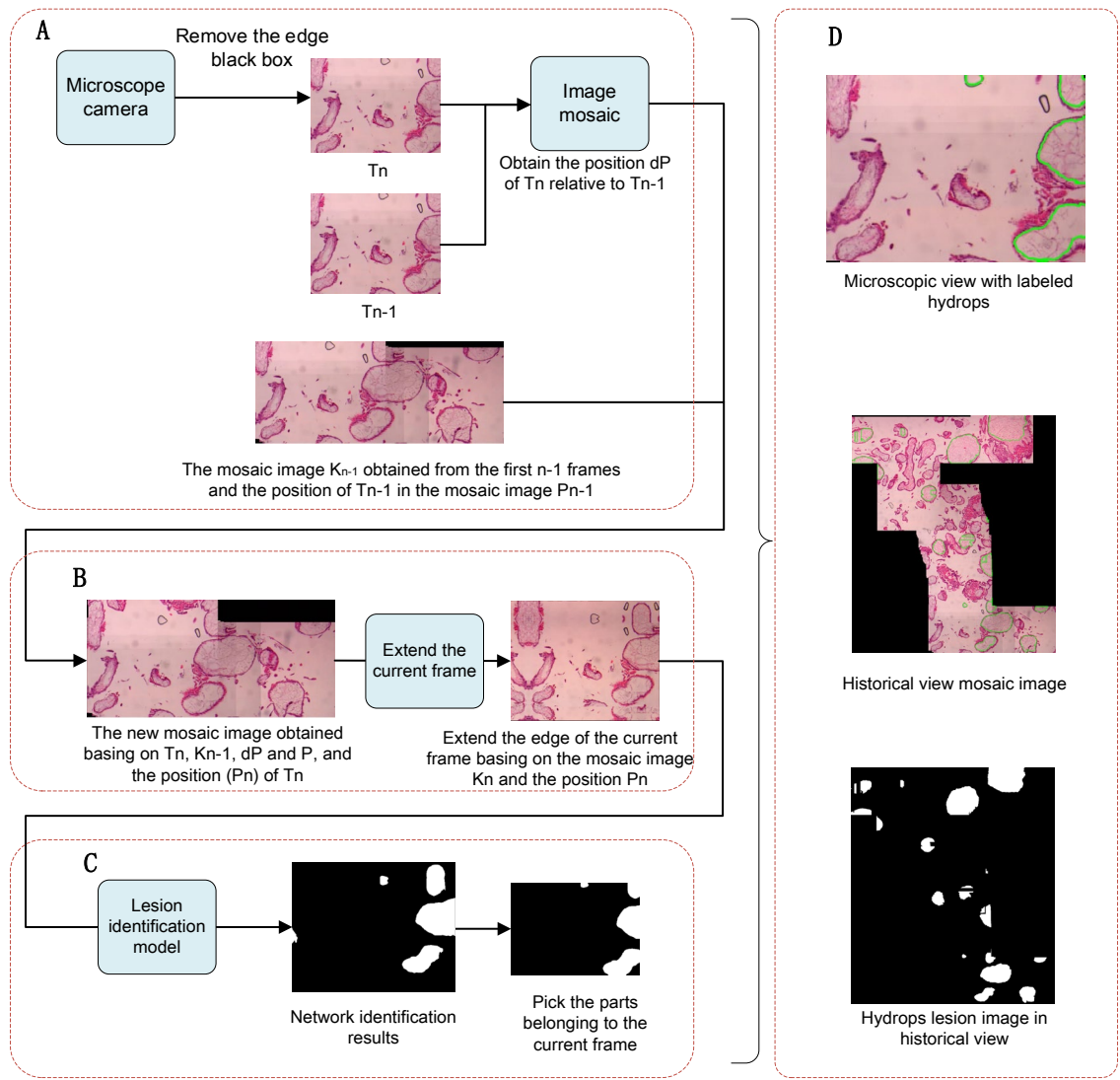

Figure 3. Workflow of the real-time CAD system.

*University* use shutter exposure. The advantage of shutter exposure is a shorter exposure time, which can reduce the motion blur on the image. However, the shutter spacing and moving speed affect the exposure effect. The downside of shutter exposure is that due to the fact that the exposure occurs region by region, the distortion and blurring of moving objects can be significant. Therefore, we need to process the collected section images. The system transmits the microscope view image sequence of two frames per second to the computer through the microscope and the microscope camera.

The workflow of the proposed real-time CAD system is shown in Figure 3, which mainly includes three modules: the image mosaic module (labeled as A in the Figure3), edge extension module (labeled as B in the Figure3), and hydrops lesion recognition module (labeled as C in the Figure3).

The image mosaic module splices the image sequence into a global HM section image, which can provide the general orientation of the current visual field in the whole section and provide doctors with more comprehensive information about the HM sections. The edge extension module expands the edge of the image in the current visual field and enriches the information of lesions by completing the edges of sections to improve the recognition accuracy. The hydrops lesion recognition model in the hydrops lesion recognition module has been proved to achieve good hydrops lesion recognition on a brand-new HM section by the experimental analysis of our unpublished work.

With the three modules, the computer can display the microscopic view, historical view mosaic, and historical view of the hydrops lesion image with hydrops lesion labels on the screen in real time.

### C. Image Mosaic Module

The microscope camera used in this paper was in shutter exposure mode. When the HM section is moved, the image will be distorted or possibly blurred.

The image mosaic module is labeled as A in Figure 3. This module is able to mosaic the current frame $T_n$ and the previous frame $T_{n-1}$ of the microscope objective view, which is obtained from the microscope camera. By mosaicking all of the historical frames, the module is able to obtain the historical view mosaic and provide the edge extension module with the edge information for extending the current frame.

Classical image mosaic algorithms are typically based on Scale-invariant feature transform (SIFT) [23] or template matching. However, this paper requires the real-time image mosaicking of HM images, so it has higher requirements for algorithm speed. In this paper, three image mosaic algorithms are selected, and the best method is obtained through comparative experiments.

The first image mosaic algorithm is Speeded Up Robust Features (SURF) [24]. SURF is an accelerated SIFT algorithm with robust local feature point detection and description. The current image is denoted as $f(x, y)$. For each pixel in the image, a Hessian matrix $H$ can be calculated. SURF determines the local maximum value through the norm of the Hessian matrix using the following equation:

$$det(H) = \frac{\partial^2}{\partial x^2} \frac{\partial^2 f}{\partial y^2} - \left(\frac{\partial^2 f}{\partial x \partial y}\right) \quad (1)$$

When the discriminant takes the maximum value, the current pixel point is the brightest or darkest point in the field of the point, so the position of the key point can be determined. In order to meet the scale invariance of images, SURF adopts box filtering before constructing the Hessian matrix, so Equation (1) is rewritten as

$$det(H) = D_{xx}D_{yy} - \left(0.9D_{xy}\right)^2 \quad (2)$$

SURF builds the scale space in the form of the pyramids. In the next step, every pixel processed by the Hessian matrix is compared with 26 points in the 2D image space and the neighborhood of the scale space to locate the feature point. The harr wavelet feature in the circular neighborhood of statistical feature points is used to match the direction of feature points and finally generate feature point descriptors. SURF determines the matching degree by calculating the Euclidean distance between two feature points. The shorter the Euclidean distance, the higher the matching degree of two feature points. A complete mosaic picture is obtained based on the matching feature descriptors in $T_n$ and $T_{n-1}$. A disadvantage is that the speed can just barely meet the requirements of real-time mosaicking of current and previous frames under the microscope. As previously mentioned, the microscope camera in this study was in shutter exposure mode. In shutter exposure mode, if the object under the microscope is moving, the image may have irregular distortion. Therefore, errors often occur in calculating feature matching, resulting in a low-quality mosaic.

The second HM image mosaic algorithm is based on affine transformation, which is the default relationship between the current and previous frame. The function we used in this study was retrieved from the OpenCV library. The advantage of this algorithm is its fast operation speed. The affine matrix is calculated by minimizing the sum of the absolute values of the difference between the gray values of the overlapping image pixels after affine transformation. This operation can be described as the convex optimization problem of equation (3).

The function $F(T_n, M)$ represents the image obtained by transforming the current frame based on affine matrix $M$. $|F(T_n, M) - T_{n-1}|$ represents the sum of the absolute values of subtracting all the corresponding pixel values of the two images.

$$min \quad |F(T_n, M) - T_{n-1}| \quad (3)$$

The affine transformation matrix M is used to transform the current frame $T_n$ and put it in the corresponding position. The mosaic process is finished by copying the previous frame $T_{n-1}$ to a specific position. However, this algorithm shares a similar disadvantage with SURF. Although it can perform fairly well when the movement of the object under the microscope is small, the affine transformation matrix cannot be calculated when the movement is larger.

We also considered a Fourier transform-based HM image mosaic algorithm. By default, the relationship between the current and previous frame is rigid translation, without stretching and deformation. If the current frame is set to be $f_1(x, y)$ and the previous frame is set to be $f_2(x, y)$, then $f_1(x, y)$ is obtained by translating $f_2(x, y)$ by $(dx, dy)$, which satisfies the following equation:

$$f_2(x, y) = f_1(x - dx, y - dy) \quad (4)$$

The two-dimensional Fourier transform of the current frame is used to obtain the frequency domain image $F_1(u, v)$, and the two-dimensional Fourier transform of the previous frame is used to obtain the frequency domain image $F_2(u, v)$, which satisfies the following equation:

$$F_2(u, v) = F_1(u, v)e^{-i \cdot 2\pi(u \cdot dx + v \cdot dy)} \quad (5)$$

The cross-power spectrum $H(u, v)$ can be obtained by multiplying and normalizing the $F_1$ image and $F_2^*$ image after conjugation.

$$H(u, v) = \frac{F_2 \cdot F_2^*}{|F_2 \cdot F_2^*|} e^{-i \cdot 2\pi(u \cdot dx + v \cdot dy)} \quad (6)$$

The real domain graph $F_e(x, y)$ is obtained by inverse Fourier transform of the cross-power spectrum $H(u, v)$. If the current and previous frames meet the rigid translation, the real domain graph $F_e(x, y)$ after inverse Fourier transform is the image of an impulse function. The displacement relation $(dx, dy)$ can be determined from the peak position in the real field graph $F_e(x, y)$ after inverse Fourier transform. According to the displacement relation $(dx, dy)$, four cases can be enumerated, namely, the previous frame is in the upper left, lower left, upper right, or lower right of the current frame, and the case of the minimum absolute value difference of gray values in the overlapping area is taken as the final mosaic result. According to the experimental results, the algorithm based on Fourier transform is slower than that based on affine transform, but it can fully meet the requirements of practical application.

### D. Edge Extension Module

The hydrops lesion recognition model has a low recognition rate for small lesions at the edge, which is caused by incomplete information. The edge extension module can expand the edge of the current visual field image and provide more comprehensive information about lesions at the edge of the visual field image.

The first edge extension scheme we adopted is setting the edge to be composed of zeros. The positions of the historical view mosaic and the current frame in the historical view mosaic are obtained through the image mosaic module. According to this information, the images of the current frame and the edge of the current frame are captured, and all the regions with no images on the edge are set to be zeros, as shown in Figure 4(a).

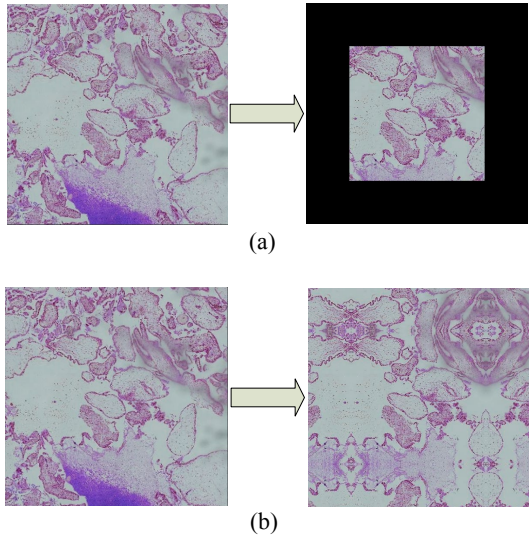

(a)

(b)

Figure 4. Examples of setting the edge extension. (a) Setting the edge to be zeros. (b) Mirror edge extension

The second scheme is mirror edge extension. This scheme also obtains the positions of the historical mosaic and the current frame in the historical view mosaic through the image mosaic module. Based on this information, the image of the current frame and the edge of the current frame is captured. The difference is that the region with the previously zeroed edge is now filled with the image of the middle region.

We integrated these two schemes into the lesion recognition model for evaluation, and the lesion evaluation metric—the intersection over union (IoU) of the middle part of the image, or the unmirrored part of the image—was finally calculated for edge images. The result of mirror edge extension is shown in Figure 4(b).

### E. Hydrops Lesion Recognition Module

The hydrops lesion recognition module is labeled as C in Figure 3. The main body of this module is the hydrops lesion recognition model. The model is able to process the input microscope view images and output corresponding labeled hydrops lesion images.

We built the HM hydrops lesion recognition model based on the semantic segmentation network (detailed information is listed in the Experiment section). First, we collected and labeled the HM hydrops lesion images and enhanced the data set as required by the hydrops lesion recognition model. Using UNet [25], Feature Pyramid Network(FPN)[26], LinkNet[27], PSPNet[28], PAN[29]and DeepLabV3+[30] networks with the backbones resnet18[31], resnet50[31], se_resnext50[32], and sernet50[32] to build the models, through multiple experiments and comparison verification, we selected the DeepLabV3+[30] model with se_resnet50 as the backbone for the lesion recognition task, as shown in Figure 5. DeepLabv3+[30] uses atrous convolution, which does not apply to the adjacent $3 \times 3$ feature convolution but to the nine feature points of the $3 \times 3$ interval rate. Atrous convolution can expand the receptive field and capture multiscale feature information. DeepLab parallels the atrous convolutions of multiple scales and combines the output results. Since the rate can be freely selected, it can adapt to different scales of lesion segmentation during network parameter adjustment, which is of great significance for the lesion segmentation of HM hydrops.

Furthermore, common loss functions such as IoULoss and BCELoss are insensitive to lesion-level evaluation metrics. Also, in clinical practice, doctors focus on different sections, so the model will require a higher recall rate. Therefore, we developed a novel compound loss function that not only takes into account pixel- and lesion-level loss but also take both recall loss and precision loss into consideration.

For a variety of evaluation metrics, we adopted a stepwise training method of the model using a variety of loss functions. The method trained the model with IoULoss or BCEWithLogitsLoss firstly, only the model with the best pixel-level IoU result on validation is saved. At the second step, the saved model is trained on the compound loss function, and only the model with the best pixel-level IoU result on validation is saved, so as to maintain pixel-level performance and improve lesion-level performance. Comparative experiments revealed that our hydrops lesion recognition model showed significant improvement in multiple evaluation indicators. This is an important step towards achieving our goal of improving the efficiency of HM diagnosis and reducing the rate of missed diagnosis and misdiagnosis. The experimental analysis is in our unpublished work.

The current frame image after edge extension was scaled to the input size of the hydrops lesion recognition model, and the processed image was input into the hydrops lesion recognition model. In the next step, the output lesion label image was scaled to the original size. After that, the corresponding edge region in the edge extension process was deleted and, finally, the labeled image of the current frame was obtained. The labeled real-time microscopic view was obtained by overlaying the edge of the labeled lesion image with the current frame image. Similar to the HM image mosaicking process, the hydrops lesion image in the historical view was mosaicked according to the position of the current frame in the historical view mosaic image, and finally we got the lesion image containing all fields.

At this point, the whole real-time system for displaying lesions in the microscopic view is able to provide physicians with the following information for effective diagnostic support: (1) Real-time microscopic view with hydrops lesion labels. (2) Historical view mosaic images, which make the physicians aware of the position of the current frame in the whole section. If the whole section is traversed, it is able to achieve a function similar to the section that scans images. (3) Hydrops lesion map in the historical view, which enables the physicians to understand the distribution of the lesion in the whole section and develop a more comprehensive understanding of the patient's condition.

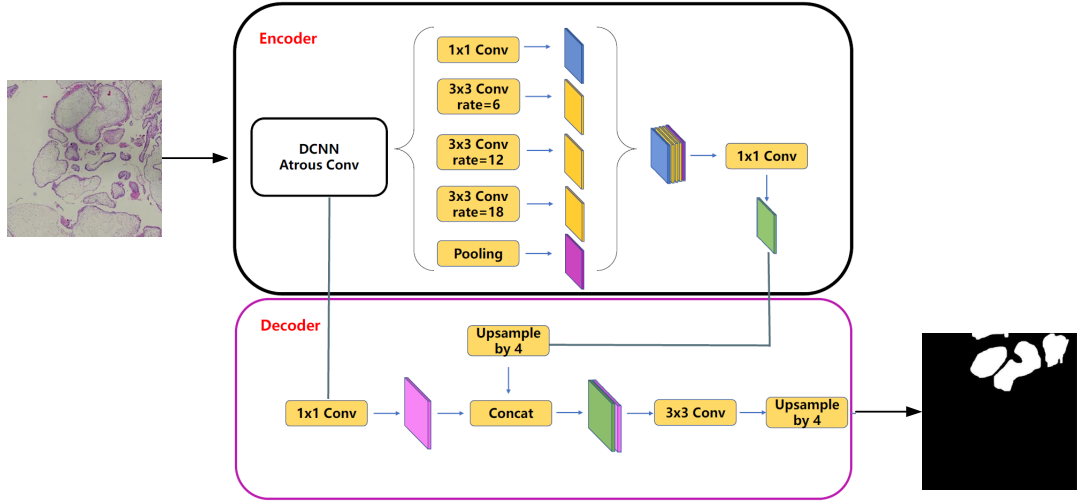

Figure 5. DeepLabv3+ network structure on HM lesion recognition.

As we can see from the preceding discussion, the three modules are generalizable to other pathological diagnostic processes as well. The purpose of the image mosaic module and the edge extension module is to improve the outcome of the hydrops lesion recognition module. The core part of the hydrops lesion recognition module is the network, which is capable of extracting the features needed to diagnose HM. As long as the network is robust, accurate, and fast, it can perform well in similar tasks after training on a relevant training set.

## III. EXPERIMENT

### A. Experimental Results of the Image Mosaic Module

In order to compare the performance of the three image mosaic algorithms in the task of mosaicking HM images under the microscope, we numbered the three algorithms as follows and then performed the experiment: M1, the HM image mosaic algorithm based on feature matching; M2, the HM image mosaic algorithm based on affine transformation; and M3, the HM image mosaic algorithm based on Fourier transformation.

Two experiments were carried out on the three algorithms, using the image set extracted from the video recorded by the microscope camera.

#### 1) Experiment 1

TABLE I
IMAGE MOSAIC ALGORITHM COMPARISON WHEN EXTRACTING ONE IMAGE
FROM EVERY FRAME

| Symbol | M1 | M2 | M3 |
|---|---|---|---|
| Time cost (ms/frame) | 198 | 62 | 124 |
| Error count | 3 | N/A | 0 |

In experiment 1, one image was extracted every frame, with 259 images in total. Image mosaicking was performed for the current and previous frames. When mosaicking, if one of the images was fixed, the second image was transformed and placed in the corresponding position based on the parameters calculated by the algorithm to get the mosaicked image. To evaluate the results, the IoU of the predicted and actual position of the second image was calculated. If the IoU was less than 0.9, then there was a mosaic error in the image. The algorithm with the least number of images with errors and the most stable mosaic process was selected as our final algorithm. The IoU calculation formula is as follows:

$$IOU = \frac{img_{true} \cap img_{pred}}{img_{true} \cup img_{pred}} \quad (7)$$

The three algorithms each extracted 258 images, and the results are listed in TABLE I.

Some of the results of experiment 1 are shown in Figure 6. We choose two groups of images with relatively large movement. The figure shows that when the movement was larger, M1 had an over-perspective-transformation problem, whereas M3, despite having a large distortion, was able to ensure an acceptable mosaic result. It can also be seen from Table 1 that this algorithm has decent robustness and no error.

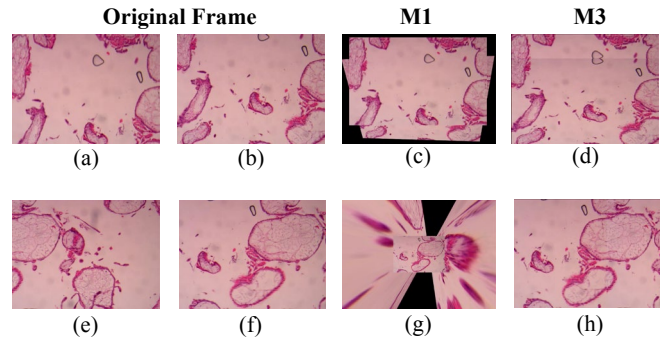

Figure 6. Results of image mosaic algorithm in experiment 1. Frames 61 and 62 are (a) and (b), frames 39 and 40 are (e) and (f). The corresponding M1 mosaic images are shown in (c) and (g), the corresponding M3 mosaic images are shown in (d) and (h).

This figure does not show the results of M2 because the affine transformation matrix of the current and previous frames cannot be calculated by the function. Moreover, M2 shares a similar limitation with M1 when the movement is large. In this case, the movement is not small enough for M2 to perform well, so the results of M2 cannot be obtained.

### 2) Experiment 2

In experiment 2, one image was extracted every five frames, with 52 images in total. Image mosaicking was performed for the current and previous frames. The three algorithms each extracted 51 images, and the results are listed in TABLE II.

TABLE II
IMAGE MOSAIC ALGORITHM COMPARISON WHEN EXTRACTING ONE IMAGE
FROM EVERY 5 FRAMES

| Symbol | M1 | M2 | M3 |
|---|---|---|---|
| Time cost (ms/frame) | 196 | 58 | 122 |
| Error count | 8 | N/A | 0 |

Some of the results of experiment 2 are shown in Figure 7. As the figure shows, when one image is extracted every five frames, the relative displacement of the current and previous frames is generally larger. The images obtained by M1 exhibited over-perspective-transformation, and the right-side villi from frame 221 was pieced together with a small portion of the leftmost edge villi from frame 226. Under the premise that the distortion of the current and previous frames cannot, in fact, be so exaggeratedly large, this algorithm is obviously inappropriate. Furthermore, M1 is prone to errors when the distinction of the feature point vector is not high, i.e., there is no global specificity of local texture features. As in experiment 1, the results for M2 cannot be obtained.

| Original Frame | M1 | M3 |

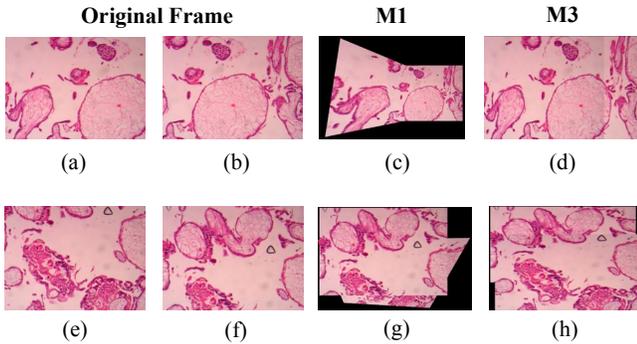

(a)     (b)     (c)     (d)

(e)     (f)     (g)     (h)

Figure 7. Results of image mosaic algorithm in experiment 1. Frames 221 and 226 are (a) and (b), frames 86 and 91 are (e) and (f). The corresponding M1 mosaic images are shown in (c) and (g), the corresponding M3 mosaic images are shown in (d) and (h).

According to TABLE I, TABLE II, and the results displayed in Figures 6 and 7, it can be seen that M1 has a relatively long calculation time and may have poor results due to the large range of perspective transformation. M2 has a relatively short calculation time but a relatively low robustness because when the movement of the current and previous fames increases, the error rate also increases. Therefore, we selected M3 as our HM image mosaic algorithm. Figure 8 shows the mosaicked historical HM image obtained by the real-time microscope camera using M3. Although there was distortion in some parts of the mosaicked image due to the shutter exposure mode of the camera, it can still meet the requirements of HM image mosaicking tasks.

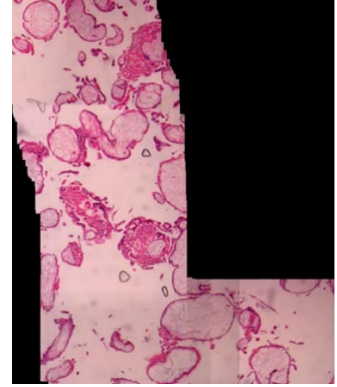

Figure 8. Mosaicked historical HM image by using M3

### B. Experimental Result of the Recognition System

The evaluation metrics of the hydrops lesions in this study included pixel-level metrics and lesion-level metrics. The pixel-level metrics were used to evaluate the recognition accuracy of the HM section in pixels. The lesion-level metrics were used to evaluate the recognition accuracy of the HM section in lesions. Introducing the lesion-level metrics takes into consideration the requirements for the clinical pathological diagnosis of HM. When clinicians make a pathological diagnosis, they mainly take a single lesion as the object of observation and diagnosis.

The pixel-level metrics and lesion-level metrics each contain three specific metrics: IoU, recall (Rec), and precision (Pre). IoU is the metric for comprehensive evaluation of the effect of lesion recognition, but physicians tend to pay more attention to Rec and Pre. Rec, which is the true positive rate—also known as sensitivity in medicine—is able to evaluate the seriousness of the missed diagnosis. Pre, which is called the positive predictive value in medicine, is able to evaluate whether the misdiagnosis is excessive. Since a missed diagnosis can have more serious consequences than misdiagnosis, Rec is more strictly required in the actual practice of pathological diagnosis.

Figure 9 shows some examples of the hydrops lesion recognition tasks completed by different networks. TABLE III shows the evaluation results of the hydrops lesion recognition tasks done by different networks on our testing dataset. We can see that DeepLabv3+ has a better comprehensive lesion recognition result than other networks. The hydrops lesion recognition effect using DeepLabV3+ after implementing the compound loss function and the stepwise training method is shown in TABLE IV. As we can see from TABLE IV, multiple evaluation metrics of this model were significantly improved because of the compound loss function and the stepwise training strategy.

In conclusion, considering the image processing time, model size, hydrops lesion recognition effect, and other factors, the recognition model developed with DeepLabV3+, the compound loss function, and stepwise training has a decent outcome in HM recognition.

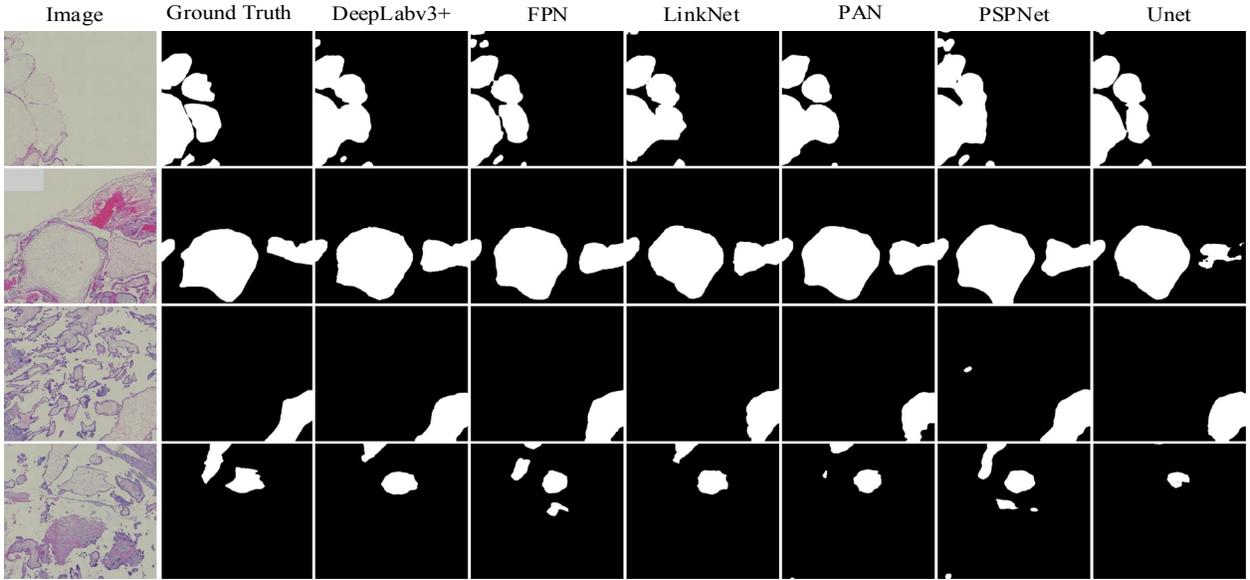

Figure 9. Examples of recognition results of different networks.

TABLE III
EVALUATION OF DIFFERENT NETWORK ON THE HYDROPS LESION
RECOGNITION TASK

| Network | PIXEL-LEVEL INDICES (%) | | | LESION-LEVEL INDICES (%) | | |
|---|---|---|---|---|---|---|
| | IoU | Rec | Pre | IoU | Rec | Pre |
| DeepLabv 3+ | 75.9 | 82.8 | **90.1** | **64.8** | 75.4 | **82.1** |
| FPN | 74.5 | 85.5 | 85.2 | 61.8 | 76.6 | 76.2 |
| LinkNet | 75.3 | 88.1 | 83.8 | 60.9 | 83.1 | 69.5 |
| PAN | 67.1 | 72.5 | 90.0 | 64.0 | 76.8 | 79.3 |
| PSPNet | 71.9 | 82.3 | 85.1 | 60.5 | **85.9** | 67.2 |
| UNet | 75.6 | **90.7** | 82.0 | 61.7 | 85.8 | 68.8 |

TABLE IV
COMPARISON OF THE WITH/WITHOUT THE COMPOUND LOSS FUNCTION AND
THE STEPWISE TRAINING

| Network | PIXEL-LEVEL INDICES (%) | | | LESION-LEVEL INDICES (%) | | |
|---|---|---|---|---|---|---|
| | IoU | Rec | Pre | IoU | Rec | Pre |
| DeepLabv3+ (Without) | 75.9 | 82.8 | 90.1 | 64.8 | 75.4 | 82.1 |
| DeepLabv3+ (With) | 77.0 | 88.1 | 94.8 | 70.2 | 86.2 | 79.1 |

## C. Experiment Result of the Edge Extension Module

Scanned section images with labeled lesions were used for validation analysis. For the sake of rigorous experiment, we first introduced two groups of comparative experiments as a baseline:

*1) Deleted edge:* delete the edge area with the width of size, which is an adjustable parameter, in all the images. The images

with deleted edges were scaled to the original image size. The images were input into the lesion recognition model, and the IoU score was recorded.

*2) Unchanged edge:* the edges of the images were left unchanged, and the original images were input into the lesion recognition model. Different from the deleted edge experiment, only the IoU score of the middle portions of the images was recorded.

On the basis of the comparative experiments, we added the edge-zero-setting and the edge-mirroring methods we introduced earlier. Unlike the edge deleting method, the edge-zero-setting and edge-mirroring methods only need to collect the IoU score of the middle portions, which are the unprocessed areas. The results are as Figure 10.

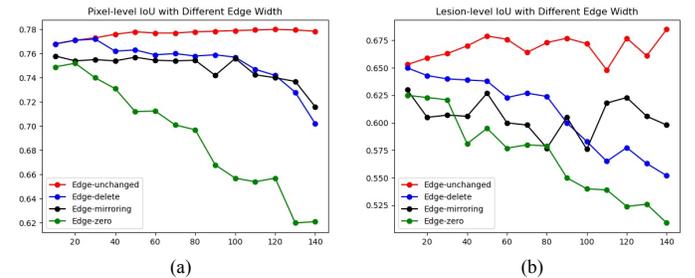

Figure 10. IoU with different edge width. (a) Pixel-level. (b) Lesion-level.

The accuracy of the unchanged edge group increased rather significantly with the increase of the edge width. However, when the edge width was around 100, the accuracy stopped increasing. This indicates that extending the edge of the images can effectively improve the overall lesion recognition rate, and the network has a rather low recognition accuracy at the edge area of the input image. For the edge-zero-setting group, both the lesion-level IoU and the pixel-level IoU decreased significantly with the edge width. This result indicates that setting the edge to be zero will have a significant impact on the hydrops lesion recognition model. Therefore, deleting the edge is not appropriate. The edge-mirroring group has serious improvement over the edge-zero-setting group. Moreover, there was little difference between the edge-mirroring group and the deleted edge group. The edge-mirroring group even has higher IoU when the edge width is over 100. This indicates that edge mirroring is able to at least not affect the recognition process of the middle portions of the images, proving the rationality of the mirroring operation in the second group.

We chose edge mirroring as our final plan, and we chose the edge width to be 120. With this setup, compared with the deleted edge group, the unchanged edge group has a 3.4% pixel-level IoU improvement and a 9% lesion-level improvement. In actual practice, the accuracy improvement of the edge extension group will mainly depend on how much edge information the historical frame mosaic images are able to provide to the current frame.

### D. Real-time CAD System Output Results

The system introduced by this paper is able to output information which can assist physicians in diagnosis. Said information is shown in Figure 11. The green boxes are identified lesions. It can be seen that basically all the hydrops lesions in the visual field can be identified effectively, and the edges of the identified hydrops lesions are also very consistent with the actual lesions.

The system has limited recognition capability for individual villi that only partially appear in the video. This will lead to missed diagnosis because the model is not able to effectively identify incomplete lesion images. However, in clinical practice, if a clinician thinks the partially appearing villi are likely to be HM villi, they will drag the visual field to the corresponding direction and continue their observation. When the whole lesion is displayed in the visual field, the network will have a much better recognition effect.

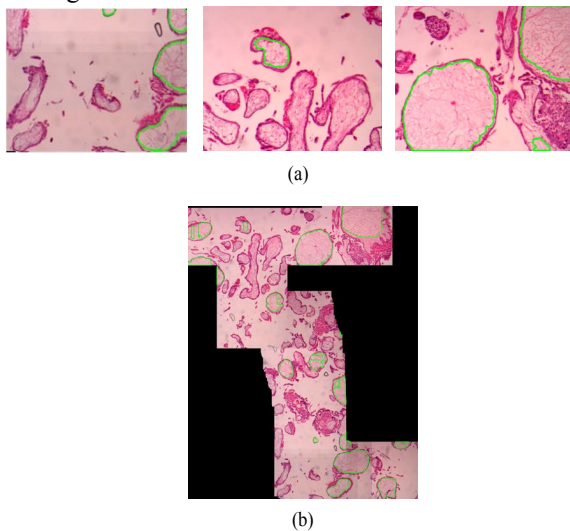

(a)

(b)

Figure 11. Real-time CAD system output. (a) Microscopic view with labeled hydrops lesion. (b) Historical vision field hydrops lesion image.

Some of the hydrops lesion areas of the villi in the historical visual field mosaic image with lesion recognition results may have faults. This is because the strategy adopted in the image mosaicking process is to cover the overlapping area of the current and previous frames with images from current frames. When there is distortion in the current and previous frames or the recognition results of the current and previous frames are different, faults will appear in the image.

### IV. CONCLUSION

In this study, a real-time CAD system for HM hydrops lesion recognition based on the microscopic view image was constructed. The system includes a microscope, microscope camera, computer host, display screen, and other hardware, as well as an image mosaic module, edge extension module, hydrops lesion recognition module, and other software modules. The system is able to provide physicians with the microscopic view, the historical visual field mosaic image, and the historical view hydrops lesion image, all with labeled hydrops lesions, in real time.

We compared and analyzed three image mosaic algorithms, and the results show that the image mosaic algorithm based on Fourier transform is the most suitable for image mosaicking under the microscope. Meanwhile, on the basis of the image mosaic module, the edge extension module was used to make the input image of the hydrops lesion recognition model have more comprehensive edge image information. The comparative experiments show that the edge extension module is able to improve the model performance by at most 3.4% regarding pixel-level IoU and 9% regarding lesion-level IoU. As for the lesion recognition model, we tested and evaluated different semantic segmentation networks and backbones, and finally we found that the DeepLabV3+ model with se_resnet50 as the backbone can most effectively identify lesions. In the recognition model, we used a compound loss function for multiple evaluation metrics, and through stepwise training of the model on multiple loss functions, the resulting model yielded significant improvements in multiple evaluation metrics. Finally, the pixel-level accuracy was 94.8% and the lesion-level recall rate was 86.2%. The real-time CAD system for HM hydrops lesion recognition from HM section images under a microscope, which utilizes a network model and is trained on clinical data, has a complete structure, clear function, remarkable effect, and high clinical application value.

Since we annotated the HM section data set to obtain a hydrops data set, hyperplasia data set, and villi data set, and because this paper mainly conducted model training on the hydrops data set, it is hoped that in the future more comprehensive and effective models can be trained on the hyperplasia data set to display hydrops and hyperplasia lesions in real time in order to better assist physicians in the diagnosis of HM.

Not only is the CAD system able to assist pathologists in the diagnosis of HM biopsy, but this system is also generalizable to other pathological practice. The hardware in the system (including the microscope, camera, and computer host) are conventional and non-specified equipment. The hardware can be used in most daily work of hospital pathology departments without any modification. Previously, we explained how the image mosaic module and edge extension module are able to improve the outcome of the hydrops lesion recognition module, and as long as mosaicking the images and extending the edges is practicable and not medically or clinically pointless, these two modules will continue to support the recognition module even if the system is used for a different pathological practice. As we discussed in the paper, the network used in the recognition module is fast enough to meet the real-time requirements. Our compound loss function and stepwise training strategy improve the robustness and accuracy of the network. The semantic segmentation network we adopted is dedicated to not only extracting features from HM sections but also extracting the corresponding features based on the training set. Despite the morphological diversity of different tissues and organs, many pathological diagnostic processes involve identifying lesions. In order to use our system to assist in diagnosing other diseases, the only modification required is to change our training data set of HM into a well-built and well-labeled training data set of the corresponding disease's pathological manifestations. In conclusion, the system we constructed in this paper is capable of facilitating not only HM diagnosis, but also other pathologic diagnoses by extracting corresponding pathological features.

## V. Acknowledgments

The authors would like to thank *the Third Affiliated Hospital of Zhengzhou University* for providing the data source.